\documentclass[conference]{IEEEtran}
\IEEEoverridecommandlockouts

\pdfoutput=1

\usepackage{cite}
\usepackage{amsmath,amssymb,amsfonts}
\usepackage{algorithmic}
\usepackage{graphicx}
\usepackage{textcomp}
\usepackage[table,xcdraw]{xcolor}
\usepackage{colortbl}
\newcommand{\grayrow}{\rowcolor[gray]{.90}}
\definecolor{lightblue}{rgb}{0.1, 0.1, 0.44}

\usepackage[urlcolor= blue]{hyperref}

\def\BibTeX{{\rm B\kern-.05em{\sc i\kern-.025em b}\kern-.08em
    T\kern-.1667em\lower.7ex\hbox{E}\kern-.125emX}}
    
\usepackage{multirow}
\usepackage{tabularx, booktabs}
\usepackage{hhline}
\usepackage{multicol}
\usepackage{enumitem}
\usepackage{lipsum}
\usepackage{amsmath}
\usepackage{flushend}
\usepackage{tikz}

\usepackage{array}

\newcolumntype{L}[1]{>{\raggedright\let\newline\\\arraybackslash\hspace{0pt}}m{#1}}
\newcolumntype{C}[1]{>{\centering\let\newline\\\arraybackslash\hspace{0pt}}m{#1}}
\newcolumntype{R}[1]{>{\raggedleft\let\newline\\\arraybackslash\hspace{0pt}}m{#1}}

\AtBeginDocument{%
  \providecommand\BibTeX{{%
    \normalfont B\kern-0.5em{\scshape i\kern-0.25em b}\kern-0.8em\TeX}}}
    
    \newenvironment{boxedtext}
    {
    \begin{center}
    
    \vspace{-4pt}
    \begin{tabular}{|p{0.96\linewidth}|}
    \hline
    }
    { 
    \\ \hline
    \end{tabular} 
    
    \end{center}
    \vspace{-3pt}
       }

\begin{document}


\title{Beyond Binary Moderation: Identifying Fine-Grained Sexist and Misogynistic Behavior on GitHub with Large Language Models}


\author{\IEEEauthorblockN{Tanni Dev}
\IEEEauthorblockA{
\textit{Department of Computer Science} \\
\textit{Wayne State University} \\
Detroit, MI, USA \\
dev.tanni@wayne.edu}
\and
\IEEEauthorblockN{ Sayma Sultana}
\IEEEauthorblockA{
\textit{Department of Computer Science} \\
\textit{Wayne State University} \\
Detroit, MI, USA \\
sayma@wayne.edu}
\and
\IEEEauthorblockN{Amiangshu Bosu}
\IEEEauthorblockA{
\textit{Department of Computer Science} \\
\textit{Wayne State University} \\
Detroit, MI, USA \\
amiangshu.bosu@wayne.edu}
}


\maketitle
\IEEEpubidadjcol

\begin{abstract}
Background: Sexist and misogynistic behavior significantly 
 hinders inclusion in technical communities like GitHub, causing developers, especially minorities, to leave due to subtle biases and microaggressions. Current moderation tools primarily rely on keyword filtering or binary classifiers, limiting their ability to detect nuanced harm effectively. 

Aims: This study introduces a fine-grained, multi-class classification framework that leverages instruction-tuned Large Language Models (LLMs) to identify twelve distinct categories of sexist and misogynistic comments on GitHub.

Method: We utilized an instruction-tuned LLM-based framework with systematic prompt refinement across 20 iterations, evaluated on 1,440 labeled GitHub comments across twelve sexism/misogyny categories. Model performances were rigorously compared using precision, recall, F1-score, and the Matthews Correlation Coefficient (MCC).
Results: Our optimized approach (GPT-4o with Prompt 19) achieved an MCC of 0.501, significantly outperforming baseline approaches. While this model had low false positives, it struggled to interpret nuanced, context-dependent sexism and misogyny reliably. 
Conclusion:  Well-designed prompts with clear definitions and structured outputs significantly improve the accuracy and interpretability of sexism detection, enabling precise and practical moderation on developer platforms like GitHub.
\end{abstract}


\section{Introduction}
\label{sec:introduction}

GitHub, the leading platform for hosting open source software (OSS) projects, facilitates extensive technical collaboration through shared repositories, code reviews, and community discussions~\cite{GitHubOctoverse2024}. Despite its critical role in supporting OSS contributions and knowledge sharing, GitHub has faced criticism for fostering hostile and exclusionary environments, particularly affecting women and other gender minorities \cite{terrell2017gender,ford2016paradise, frluckaj2024codes}. 
Research indicates that women on GitHub often experience implicit biases~\cite{sultana-EMSE-2023}, dismissive behaviors~\cite{singh2022discrimination}, and overt discrimination~\cite{terrell2017gender}, resulting in disproportionately low participation~\cite{bosu2019diversity} and retention rates~\cite{almukhambetova2021fixing,ashcraft2016women}.

To address the underrepresentation and improve the retention of gender minorities, researchers have proposed several strategies to combat sexism and misogyny within software development communities. These efforts include adopting Codes of Conduct (CoC)~\cite{singh2022codes,li2021code,tourani2017code}, implementing Implicit Bias Training (IBT)~\cite{dagan-google,matthiesen2023implicit}, and developing automated tools to detect harmful communication~\cite{sultana2023automated,sultana2024exploring,sarker-tosem-toxicr,ferreira2024incivility}. Despite their potentials, each strategy encounters notable limitations. Enforcing CoCs, for instance, can be resource-prohibitive in large OSS projects due to the sheer volume of interactions~\cite{singh2022codes,sarker-tosem-toxicr}. Similarly, while IBTs aim to alter individual behaviors, studies indicate their impact is often short-lived without sustained structural support and broader organizational interventions~\cite{forscher2019meta,lai2013reducing}.

Automated tools, often suggested as aids for CoC enforcement, present their own set of challenges. A primary issue is their limited contextual understanding; as current tools~\cite{sarker2023toxispanse,sarker-tosem-toxicr,ferreira2024incivility} rely heavily on the presence of a certain set of keywords for making decisions, which hinders their ability to identify subtle, context-dependent, or nuanced forms of harmful language, particularly on platforms like GitHub. 
Moreover, sexism and misogyny manifest in varied forms, such as discrediting remarks, sexual harassment, stereotyping, maternal insults, and objectification~\cite{sultana2021rubric}, each with distinct implications that demand tailored moderation approaches. Current tools, however, often rely on binary classifications, which fall short of the granularity required for such differentiated analysis. Compounding this, their lack of interpretability poses a further obstacle, as maintainers struggle to comprehend or trust the tools’ decision-making processes, especially when navigating contextual or cultural variations~\cite{sultana2023automated,vidgen2019challenges}.

To address this gap, we aim \textit{to develop an explainable multi-class classifier designed to identify diverse forms of sexist and misogynistic texts within GitHub communications.} 
Our tool is built around carefully designed and iteratively refined prompts that are tailored for state-of-the-art (SOTA) Large Language Models (LLMs). These prompts include contextual cues to improve the detection of subtle, platform-specific expressions of sexism and misogyny, particularly those commonly found on GitHub. A key feature of our design is that the prompts guide the LLMs to perform multiclass classification, enabling differentiation among various types of sexism and misogyny. Additionally, the tool generates justifications for each classification decision, thereby enhancing interpretability and transparency.

Unlike traditional supervised classifiers, which require large, manually annotated datasets for training, our approach bypasses this significant bottleneck. Constructing such datasets is often prohibitively resource-intensive due to the laborious nature of annotation and the relative scarcity of some forms of sexist and misogynistic content~\cite{sultana2023automated}. By leveraging LLMs, we offer a scalable and efficient alternative that supports nuanced multiclass classification without the need for exhaustive manual data preparation. Moreover, results of recent studies also show that instruction-tuned LLMs outperform conventional classifiers in multi-class text classification tasks~\cite{zhao2024advancing,kostina2025large}.

We evaluated our approach by applying three SOTA LLMs to a dataset of 1,440 GitHub comments, which were manually annotated by multiple annotators using a 12-category classification scheme. To evaluate the effectiveness of our prompts, we employed standard metrics—precision, recall, F1-score, and the Matthews Correlation Coefficient (MCC)—providing a comprehensive measure of performance across imbalanced classes. Through an analysis of misclassifications and the justifications accompanying them, we iteratively optimized our prompts. The best-performing version achieved an MCC of 0.501 and 84.1\% accuracy. It also achieves  98.25\%  binary precision (i.e., marks a text as sexist or misogynistic if it belongs to one of its 11 forms), 76.6\% binary recall, and 86.1\% binary F1-score. We also identified key lessons learned that can inform the development of domain-specific multiclass text classifiers tailored to Software Engineering (SE) contexts, offering practical insights for future research.

 \vspace{2pt}
\noindent \emph{Key contributions:} 
\begin{itemize}
\item First multiclass classifier designed to categorize sexism and misogyny in GitHub communications.
\item Comparative analysis of three SOTA LLMs for detecting sexism and misogyny in a SE context.
\item Empirical evaluation driving iterative prompt optimization for multiclass text classification.
\item Curated lessons for developing SE-specific multiclass classifiers leveraging LLMs.
\item We have \href{https://github.com/WSU-SEAL/github_comment_classification} {publicly released} prompts, evaluation results, and dataset for broader access and replication.
\end{itemize}

 \emph{Organization:} The remainder of this paper is organized as follows. 
  Section~\ref{sec:research-method} details our research method. 
 Section~\ref{sec:results} presents our evaluation results. 
 Section~\ref{sec:discussion} and \ref{sec:threats-to-validity} discuss lessons learnt and threats to validity, respectively.
 Section~\ref{sec:background} provides an overview of the related works.
Finally, Section~\ref{sec:conclusion} concludes the paper.

\section{Research Method}
\label{sec:research-method}

We adopt the Design Science (DS) Research methodology~\cite{dresch2014design}. Following the five stages of the DS cycle, we started with the practical problem of accurately classifying sexism and misogyny. In the design stage, prepare a test dataset that ensures balanced representation and construct a prompt containing task instructions, behavior-based category definitions, and real GitHub examples to improve classification accuracy and interpretability. In the evaluation stage, we assess prompts and LLMs using precision, recall, F1-score, and the Matthews Correlation Coefficient (MCC) to compare models and categories related to sexism and misogyny. We also iteratively refined the prompts based on our analysis of misclassification (reflection and refinement). Finally, we use this understanding to iteratively refine the prompts, and this cycle repeats (redesign). 
\subsection{Sexism and Misogyny}
In this context, our definitions of sexism and misogyny are as follows. 
\textit{Sexism is discrimination or prejudice based on a person's sex or gender, whereas misogyny is the hatred or ingrained prejudice specifically against women.}

\subsection{Dataset selection}
We use the publicly available Sexual Orientation and Gender Identity-based Discrimination (SGID) dataset from Sultana \textit{et} al.~\cite{sultana2023automated}. This dataset comprises 1,422 GitHub issue comments belonging to one of the 12 SGID categories, as well as 9,585 non-SGID comments. We select this dataset, since \textcircled{\raisebox{-0.9pt}{1}} it is the only publicly available dataset including real-world examples of sexism and misogyny on GitHub; \textcircled{\raisebox{-0.9pt}{2}} it was manually labeled using a diverse group of human annotators; \textcircled{\raisebox{-0.9pt}{3}} the authors reported a reliable inter-rater reliability for their multiclass labeling; and \textcircled{\raisebox{-0.9pt}{4}} 11 out of their 13 categories can be mapped to a sexism or misogynistic forms. 

To clarify the label definitions of this dataset, we briefly outline the original annotation methodology~\cite{sultana2023automated}. The dataset comprises GitHub issue comments, sourced via GHTorrent/API, and manually annotated by three researchers with domain expertise in software engineering and bias detection. Using Label Studio, annotation occurred over three iterative rounds. This rigorous process involved independent labeling followed by discussion and rubric refinement to resolve disagreements, ensuring careful consideration of tone, intent, and context. The labeling was grounded in a behaviorally defined schema, adapted from prior misogyny detection rubrics~\cite{sultana2021rubric} and expanded to capture LGBTQ+ harm and developer-specific nuances, accounting for both explicit and implicit forms of harm within technical discourse. Inter-annotator reliability improved across iterations; initial agreement yielded Fleiss' Kappa values of $\kappa=0.658$ (binary) and $\kappa=0.421$ (multi-label classification), increasing to $\kappa=0.692$ in the final round. While acknowledging the inherent subjectivity in manual labeling, the original study mitigated this through this iterative refinement process, structured team discussions, and clear documentation~\cite{sultana2023automated}.

\subsection{Dataset Preparation}
Sultana \textit{et} al.'s SGID dataset exhibits significant imbalance, with categories such as \emph{Threats of Violence, Damning, and Dominance} being underrepresented. To create a balanced dataset suitable for our experiment, we established a target of 40 examples per category. This target was selected pragmatically, as five categories in the original dataset already met or exceeded this number. For these categories with sufficient data, we randomly sampled 40 examples if more were present. We excluded \emph{sexual reference} and \emph{Others} categories as those do not fit our definition of sexism or misogyny. We focused on augmenting the remaining six underrepresented ones. To achieve this, we utilized GitHub's text search feature with keywords identified by Sultana \textit{et} al.~\cite{sultana2023automated} to find relevant issue comments. Potential comments were shortlisted by one author and subsequently verified by a second; comments were excluded upon disagreement to ensure reliability. This process yielded 146 new samples across the six targeted categories. Furthermore, to better reflect real-world data distributions where neutral comments are prevalent and to reduce the risk of models overfitting to positive examples, we randomly selected 1,000 samples labeled as 'Neutral' by Sultana \textit{et} al.~\cite{sultana2023automated}. Table~\ref{tab:category_distribution_final} details the category definitions~\cite{sultana2023automated}, provides examples from our final dataset, and shows the sample distribution across the original and newly collected sources.

\begin{table*}[h]
\centering
\caption{Category Definitions and Distribution in Final Evaluation Dataset. The column `Sultana \textit{et} al.' shows the number of samples taken from the SGID dataset. The column `New' shows samples searched and labeled by us.}
\label{tab:category_distribution_final}
\resizebox{\linewidth}{!}{
\begin{tabular}{p{2cm}|p{5cm}|p{5cm}|R{1.8cm}|R{0.9cm}|}
\hline
\textbf{Category } & \textbf{Definition} & \textbf{Example} &\textbf{ Sultana \textit{et} al.} & \textbf{New} \\
\hline

Discredit & Criticizes someone’s skills or intelligence in a disrespectful or dismissive way. & “Seriously, do you even know how pointers work? Let someone competent review this.” & 40 &0  \\
\grayrow
Stereotyping & Makes a general statement based on gender roles or traits. & “Figures—it’s probably written by a woman, so it’s all comments and no real code.” & 24 & 16 \\

Sexual Harassment & Aggressive or mocking language about someone’s sexuality or sexual behavior. & “Why don’t you take a break from coding and come sit on my lap instead?” & 23 &7 \\
\grayrow
Threats of Violence & Expresses intent to harm someone through physical or violent threats. & 	“Push one more buggy commit and I’ll come over there and smash your laptop.” & 16& 24 \\

Maternal Insults & Targets someone by insulting their mother or other female relatives. & “Your mom could’ve written cleaner code in her sleep.” & 40 & 0\\
\grayrow
Sexual Objectification & Reduces a person to a sexual role, body part, or object. & “Nice PR, but honestly I’m more interested in those curves than your code.” & 40 &0  \\

Anti-LGBTQ+ & Uses slurs or hostile language against LGBTQ+ identities. & “Keep that gay agenda out of this repo—we’re here to ship code, not propaganda.” & 40 & 0 \\
\grayrow
Physical Appearance & Comments on someone’s looks, body, or clothing in a critical way. & “Maybe fix your hair after you fix these bugs; it’s a mess in both cases.” & 40  & 0 \\

Dominance & Attempts to control, silence, or push women out of the conversation. & “Step aside—real engineers are taking charge now, and you can watch from the sidelines.” & 11  & 29\\
\grayrow
Damning & Expresses strong hate, shame, or moral condemnation. & “You’re an embarrassment to this project; just delete your account and disappear.” & 9&31 \\

Victim Blaming & Blames the person who experienced harm instead of the one who caused it. & 	“If you can’t handle a little harassment, maybe open source isn’t for you.” & 1&39 \\
\grayrow
None & Neutral or technical comment with no harmful content. & “Let’s refactor this function to lower its cyclomatic complexity.” & 1,000 &0 \\

\hline
\end{tabular}
}
\vspace{-10pt}
\end{table*}

\subsection{Data Cleaning} 
Before submitting the data for LLM classification, we performed standard text preprocessing to ensure consistency and mitigate variability introduced by different LLM-internal routines. This involved punctuation normalization, whitespace collapsing, and the removal of special characters.

\begin{figure}[h]
\centering
\includegraphics[width=0.9\linewidth, trim={0 0.9cm 0 0},clip]{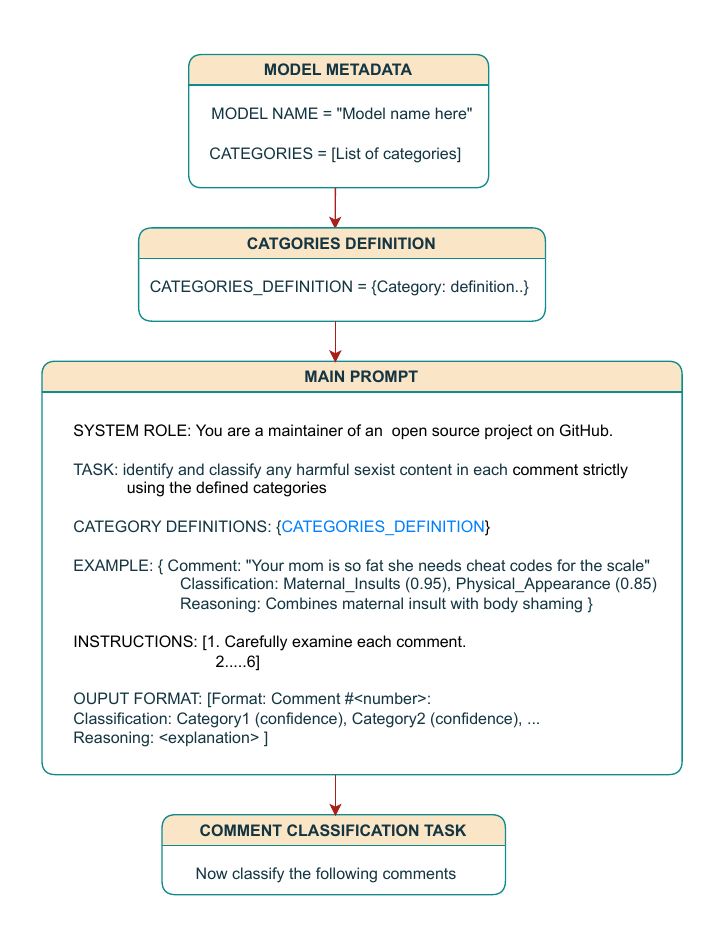}
\caption{Common Structure of our LLM prompts}
\vspace{-15pt}
\label{fig:prompt_design}
\end{figure}

\subsection{Prompt Design}
Figure~\ref{fig:prompt_design} illustrates the overall structure of the prompts used in this study, comprising four key sections: model metadata, classification schema definitions, the main prompt body, and the specific classification task input.
The main prompt section, which underwent optimization as detailed in Section~\ref{sec:prompt-optimize}, contained the following six distinct components designed to guide LLMs effectively.

\begin{enumerate}
\item \emph{Role Clarification:} Instructed the model to adopt the following persona to frame its responses appropriately. {\textit{``You are a maintainer of an open source project on GitHub. You aim to maintain an inclusive environment for persons representing various gender identities.''}} 

\item \emph{Task Assignment:}  Directed LLMs {\textit{``to identify and classify any sexist or misogynistic content in each comment to one of our predefined categories''}}, defining the core objective.

\item \emph{Category Definitions:} Provided concise, iteratively refined definitions for each category, informed by prior research~\cite{sultana2021rubric, anzovino2018twitter}, to ensure clarity and consistent model interpretation.

\item \emph{Examples (Few-Shot Learning):} Incorporated one or more real-world GitHub examples, each labeled with appropriate categories, confidence scores, and brief reasoning to align the model with the specific task using few-shot learning. The number and content of these examples were varied during prompt optimization (Section~\ref{sec:prompt-optimize}).

\item \emph{Classification Guidelines:} Included explicit instructions on handling comments fitting multiple categories, assigning confidence scores (0.00-1.00), providing brief reasoning ($\leq$ 20 words), and adhering strictly to the predefined output format.

\item \emph{Output Format Specification:} We provided output formatting instructions to standardize evaluation. When no harmful content was detected, the model was instructed to return the \emph{Neutral} label with a high confidence score {($\geq 0.95$)}. For sexist or misogynistic content, we asked the model to classify it into one or more relevant labels, along with confidence scores and a brief explanation.
\end{enumerate}

This initial version, referred to as the {Prompt 00}, served as a baseline for iterative improvements.

\subsection{Prompt Refinement}
\label{sec:prompt-optimize}
\begin{figure*}[t]
\centering
\includegraphics[width=0.8\linewidth]{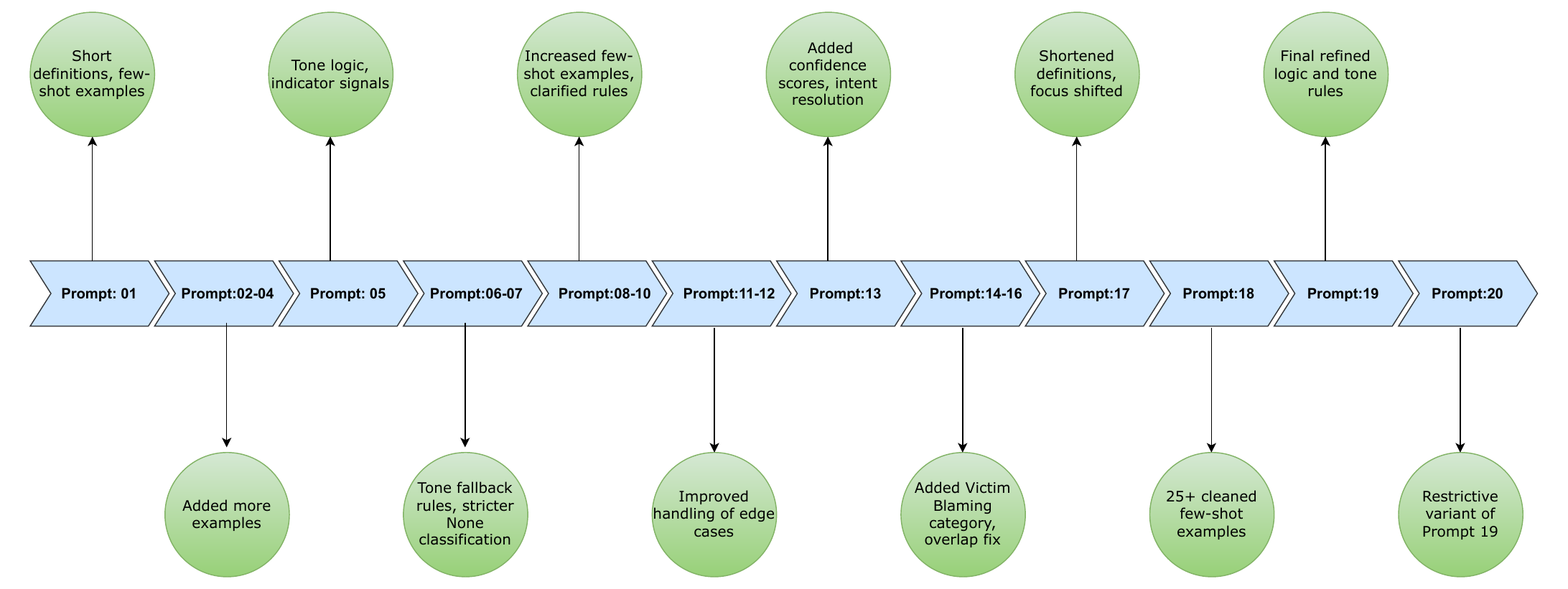}
\vspace{-10pt}
\caption{Iterative prompt-design structure highlighting the key refinements introduced at each version}
\label{fig:prompt_design_technique}
\vspace{-10pt}
\end{figure*}

The prompt underwent 20 iterations to improve classification performance through error analysis (Figure~\ref{fig:prompt_design_technique}). Initial versions had brief definitions and few examples, causing high false positives and confusion among related sexism/misogyny categories. Key refinements were:
\textcircled{\raisebox{-0.9pt}{1}} Expansion of category definitions with behavior-based cues (e.g., dismissive tones, stereotypes, subtle competence attacks).
\textcircled{\raisebox{-0.9pt}{2}} Inclusion of few-shot examples (2--3 per category by Prompt 19) for diverse, sarcastic, and overlapping harms, e.g., sarcastic comments misclassified as harmless. Examples were carefully selected to avoid noise or overfitting, with a focus on weak categories such as \emph{Objectification}, \emph{Maternal Insults}, and \emph{Physical Appearance}.
\textcircled{\raisebox{-0.9pt}{3}} Inclusion of explicit rules to resolve overlapping categories (e.g., prioritizing dominant intent in \emph{Discredit} vs. \emph{Dismissing}) and handle multi-intent or sarcastic language.

To isolate prompt design effects, we used a single inference run per prompt-model pair with fixed parameters (\emph{temperature = 0.1}, \emph{top-p = 0.9}). Deterministic inference ensured consistent outputs, making one run sufficient (Figure~\ref{fig:result-evolve}). Model reasoning outputs guided refinements by revealing misjudgments, e.g., missing sarcasm or stereotypes.

For example, the comment \textcolor{teal}{``\textit{I bet Karen’s clients will get a kick out of these...}''} (gold label: \textbf{Stereotyping} due to ``Karen’’) was misclassified as \textit{Neutral} under Prompt 14. Prompt 19 addressed this by: (1) expanding \emph{Stereotyping} to include gendered tropes, (2) adding a rule to classify such references, and (3) including a similar example (e.g., “What does a blonde do...”). This led to correct classification across GPT-4o, Mistral 7B, and LLaMA 3.3, with GPT-4o reasoning: \textit{“Uses ‘Karen’ as a gendered trope.”}

Prompt 19, with 33 diverse examples and rules for tone and overlap, achieved optimal performance but risked overfitting. We mitigated this with limited, varied examples to ensure generalization.

\subsection{Parameter Tuning}
To ensure stable and interpretable model outputs, we tuned generation parameters: low temperature (\emph{temperature = 0.1}) for consistency, high top-p (\emph{top-p = 0.9}) for diversity without incoherence, and low maximum token limit (\emph{max tokens = 150}) for concise responses. Early tests with higher temperatures (e.g., 0.7) showed inconsistent outputs, which were addressed through comparisons between deterministic (temperature = 0) and low-randomness settings. This configuration balanced stability and expressiveness, reduced variation, and was applied uniformly across models to ensure fair performance comparisons.

\subsection{LLMs Evaluated}
We conducted a comparative evaluation of various instruction-tuned large language models (LLMs) to assess their effectiveness in handling fine-grained, multi-class classification tasks within software engineering contexts. The models evaluated include both proprietary and open-source options.

\begin{table}[htbp]
\centering
\caption{Language Language Models Used in this Study}
\label{tab:languge-model-list}
\resizebox{\linewidth}{!}{
\begin{tabular}{|p{1.7cm}|p{0.5cm}|p{1.5cm}|p{1.5cm}|p{1.4cm}|p{1.1cm}|}
\hline
\textbf{Model} & \textbf{Size} & \textbf{Provider} & \textbf{Access Method} & \textbf{Release } & MCC (Phase 1) \\
\hline
GPT-4o & — & OpenAI & OpenAI API & May  2024 & 0.253 \\
GPT-3.5 Turbo & — & OpenAI & OpenAI API & March  2023 &0.156 \\
LLaMA 3.3  & 70B & Meta & TogetherAI & Dec  2024 & 0.235 \\
Mistral 7B & 7B & Mistral AI & TogetherAI & Sep  2023 & 0.171\\
DeepSeek V3 & — & DeepSeek & TogetherAI & Dec  2024 &   0.165\\
DeepSeek R1 & — & DeepSeek & TogetherAI & Jan  2025 &0.134 \\
\hline
\end{tabular}
}
\vspace{-5pt}
\end{table}

Although all six models were examined during our Phase 1, only the three best-performing ones—GPT-4o, Mistral 7B, and LLaMA 3.3—were selected for detailed Phase 2 evaluation due to their superior accuracy, consistency, and interpretability.

\subsection{Evaluation Design}
The evaluation had two phases. In Phase 1, we used a dataset of 320 comments (100 neutral, 20 per 11 sexism/misogyny categories) to assess six LLMs (Table~\ref{tab:languge-model-list}). GPT-4o, LLaMA 3.3, and Mistral 7B outperformed others based on MCC scores, while DeepSeek struggled with nuanced comments (e.g., sarcasm, workplace tones).

In Phase 2, we used the full 1,440-comment dataset and the top three models. We tested prompt configurations in three stages: zero-shot (Prompt 00, category definitions only), one-shot (Prompt 01, one example with label and reasoning), and refined few-shot prompts (Prompts 02--19, with enhanced tone cues, overlap resolution, and example diversity; Table~\ref{tab:prompt_evolution}). GPT-4o’s MCC improved from 0.349 (Prompt 00) to 0.367 (Prompt 01) and further with refinements (Table~\ref{tab:mcc-compare}). This iterative process targeted weaknesses like category confusion and sarcasm detection, ensuring consistent model and generation settings to evaluate prompt structure impact on accuracy and robustness.

\begin{enumerate}
\item \textit{Precision}: Proportion of correct positive predictions per category.
\item \textit{Recall}: Proportion of actual positives correctly identified per category.
\item \textit{F1-score}: Harmonic mean of precision and recall.
\item \emph{Matthews Correlation Coefficient (MCC)}: Primary metric for imbalanced multi-class settings. MCC incorporates all confusion matrix components, yielding a value from -1 to +1 (+1 for perfect classification, 0 for random guessing, and negative for systematic errors). It’s ideal for skewed distributions (e.g., \textit{Neutral} class dominance) where precision, recall, or macro-averaged F1 may mislead. MCC provides a single, interpretable score for global performance. For example, GPT-4o’s MCC improved from 0.349 (zero-shot, Prompt 00) to 0.501 (Prompt 19), showing gains in underrepresented classes like \emph{Dismissing}, \emph{Discredit}, and \emph{Sexual Objectification}.

\end{enumerate}
MCC’s comprehensive nature made it key for assessing robustness and progress in prompt refinement.

\section{Evaluation Results}
\label{sec:results}
The following subsections detail our evaluation results, which answer our four evaluation questions. 

\subsection{How does each model perform with our baseline zero-shot prompt?}

\begin{figure*}[h]
    \centering
    \includegraphics[width=0.90\textwidth]{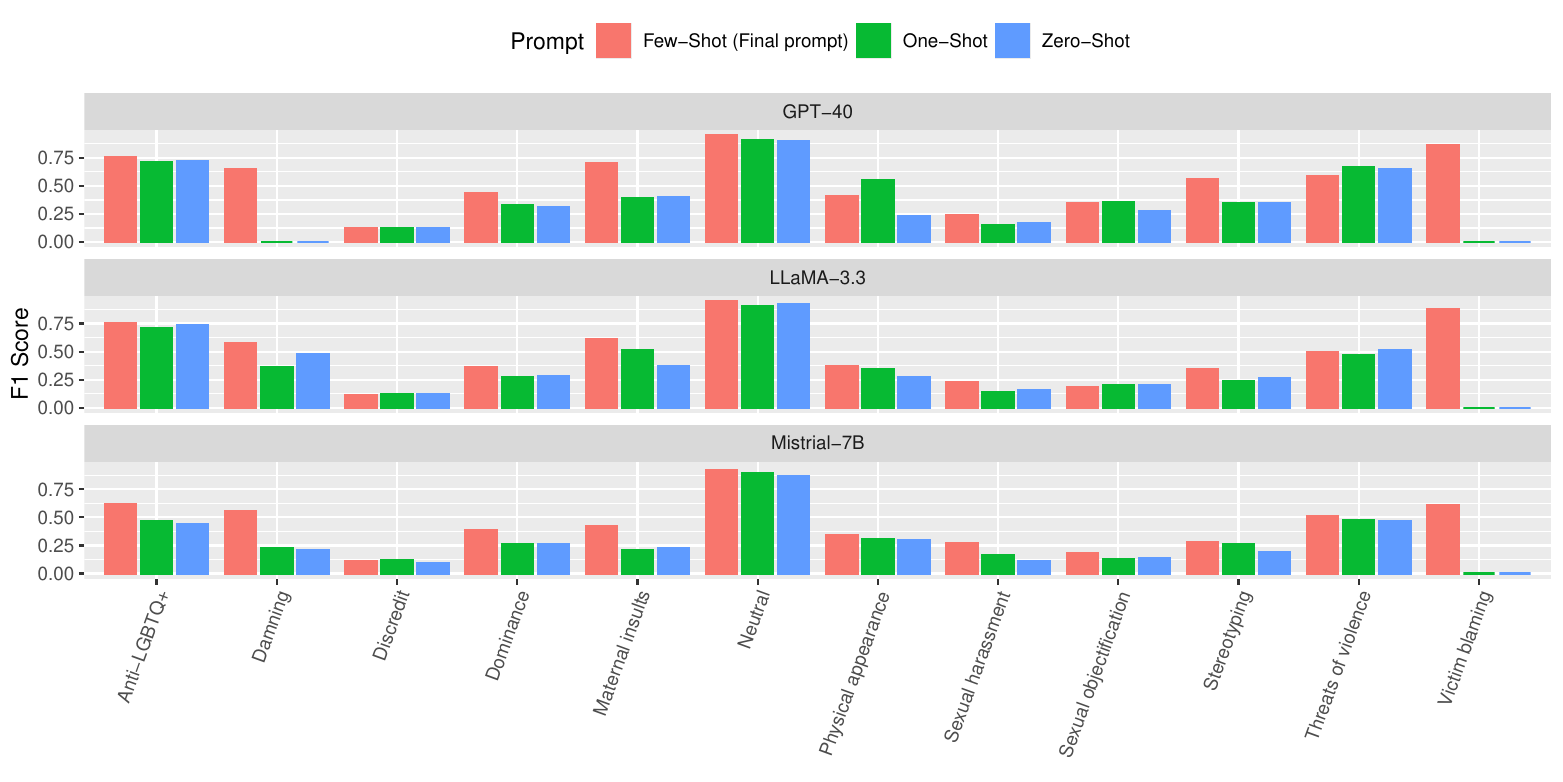}
    \caption{ Comparison of F1-scores based on Zero-Shot, One-Shot, and Few-Shot prompts}
    \label{fig:result-evolve}
    \vspace{-15pt}
\end{figure*}

\begin{table}
    \centering
    \caption{MCC scores of the best three models with zero-shot, one-shot, and refined few-shot prompts}
    \label{tab:mcc-compare}
    \begin{tabular}{|l|r|r|r|}
    \hline
        \emph{Prompt} &\emph{ GPT-4o }& \emph{LLaMA-3.3} & \emph{Mistrial-7B}  \\
        \hline
        Zero-shot & 0.349 & 0.354 & 0.270 \\
        One-Shot & 0.367 & 0.343 & 0.306 \\
        Few-shot (refined) & 0.501 & 0.451 & 0.362 \\ 
        \hline
    \end{tabular}
    \vspace{-10pt}
    
\end{table}

We started with the first prompt (\emph{Prompt 00}), which provided only brief category definitions from Sutana et al.'s paper, no examples (zero-shot), and minimal structure.  Llama 3 achieved the highest overall MCC score with this prompt (Table~\ref{tab:mcc-compare}, first row). However, category-specific performance revealed significant limitations. As shown by F1-scores (Figure~\ref{fig:result-evolve}, blue bars), all models performed well on the \emph{Neutral} category, indicating low false positives. However, performance was consistently poor for several sexism and misogyny categories, particularly \emph{Victim blaming}, \emph{Damning},  and \emph{Discredit}. 

This baseline performance demonstrated that all models struggled in detecting subtle harms in zero-shot settings. Categories that occur more frequently in real-world (e.g., \textit{Neutral}) achieved better scores, whereas those requiring more nuanced interpretation and rare occurrence exhibited lower F1-scores.

\begin{boxedtext}
    \textit{Finding \#1:} LLaMA 3.3 performed the best with our initial zero-shot prompt. While all models performed well for the \textit{Neutral} class, they struggled with samples from \emph{Victim blaming}, \emph{Damning},  and \emph{Discredit} categories.
\end{boxedtext}

\subsection{How does model performance evolve during prompt refinement?}

Table~\ref{tab:prompt_evolution} summarizes the progression of prompt versions, potential performance-limiting factors,  key improvements, and observed performance trends.
An early insight during prompt refinement was that abstract category names hindered model comprehension. The initial \emph{Victim Blaming} label exemplified this. In Prompt 14, we replaced it with \emph{Deflection}, focusing on clearer behavioral cues, such as blame-shifting or downplaying events. While this significantly improved GPT-4o's performance (Precision=0.95, F1-score=0.75), LLaMA and Mistral remained unable to classify any examples (all scores 0.00). Subsequently, we further refined the label to \emph{Dismissing}—a broader term encompassing invalidation, minimization, and disregard. This revised label, combined with improved examples, enabled all three models to successfully identify instances of this category for the first time. 

\begin{boxedtext}
    \textit{Finding \#2:} Label phrasing significantly impacts model comprehension. Abstract terms like \emph{Victim Blaming} led to confusion, while behaviorally descriptive labels such as \emph{Deflection} and the later refinement to \emph{Dismissing} improved model performance. 
\end{boxedtext}
\vspace{4pt}

As shown in Table~\ref{tab:prompt_evolution}, performance improved consistently across subsequent iterations, culminating in Prompt 10, which achieved the highest MCC scores overall (MCC=0.501), with GPT-4o. 
Compared to the Prompt 00, Prompt 19 incorporated several key refinements: (1) confusing or abstract categories were renamed (e.g., \emph{Victim Blaming} evolved to \emph{Dismissing}); (2) 33 carefully selected examples covering edge cases like sarcasm, overlapping intent, and indirect language were included;  (3) decision rules were added to handle multi-label ambiguity and tone-based misclassification, and explicit instructions to consider dominant intent, sarcasm, and tone.
With Prompt 19, the \emph{Victim Blaming} category, which had 0 F1-score with Prompt 00, achieved strong F1-scores across models (GPT-4o: 0.86, LLaMA: 0.88, Mistral: 0.61), demonstrating the substantial impact of iterative label clarification and example optimization.

To further demonstrate this improvement, consider two comments that were misclassified in earlier prompts but correctly identified after refinement. The comment \textcolor{teal}{``\textit{``If you had an issue why didn’t you bring it up sooner''}} was consistently labeled as \emph{Neutral} in Prompt 13 by all three models, despite implying fault on the victim. Prompt 14 introduced the renamed label \emph{Deflection},with expanded definition to highlightblame-shifting and credibility undermining. However, few-shot examples were still sparse. GPT-4o again labeled the comment Neutral, with slightly improved reasoning; LLaMA and Mistral also failed. Only after the renaming to \emph{Dismissing} in Prompt 19---with a more structured definition covering minimization, deflection, silencing, and blame-shifting. New examples aligned with these patterns and targeted classification guidelines were added. This time, both GPT-4o and Mistral correctly labeled the comment as Dismissing, citing its tendency to shift blame onto the victim. Only LLaMA continued to misclassify.

Another example, \textcolor{teal}{``\textit{``I get that you feel uncomfortable but are you sure it wasn’t just a misunderstanding?''}}, was also misclassified by all models under Prompt 13.  In Prompt 14, only GPT-4o correctly identified it as \emph{Dismissing}. By Prompt 19, all three models aligned with the gold label.

These examples show how targeted revisions—such as renaming categories, expanding behavioral cues, and refining examples—enabled better detection of subtle harms that earlier prompts failed to capture.

\begin{table*}
\centering
\caption{Iterative refinement and evaluation from Priompt 00 to 19 with MCC scores}
\label{tab:prompt_evolution}

{
\centering
\resizebox{\linewidth}{!}{
\begin{tabular}{|p{1cm}|p{7cm}|p{4.5cm}|p{1cm}|p{1cm}|p{1cm}|}
\hline
\textbf{\#} & \textbf{Iterative Improvements} & \textbf{Issues Root cause → outcome} & \textbf{GPT4o} & \textbf{LLaMa} & \textbf{Mistral} \\
\hline
\grayrow
00 & Zero-shot prompt, with each category briefly described. & Low performance among several classes & 0.349 & 0.354 & 0.270 \\
01 & Each category is defined in a single, direct sentence; eight total few-shot examples provided with confidence score and short reasoning. &  Minimal guidance → High false-positive rate and blurry category boundaries. & 0.350 & 0.323 & 0.282 \\
\grayrow
02–04 & Definitions expanded to two sentences; four additional examples added. & Limited context despite added examples → false-positive rate still high, though boundary clarity improves.  & 0.364 & 0.361 & 0.340 \\
05 &  Added guidance on tone and sarcasm and behavioral “red-flag” cues; the examples expanded to twelve.    & Absence of tone / sarcasm cues → Misclassified sarcasm reduced after cues introduced. & 0.330 & 0.366 & 0.325 \\
\grayrow
06–07 & Example set kept roughly the same; introduced a conservative fallback that labels ambiguous text as harmful and narrowed the criteria for the None category. & Ambiguous comments labeled “None” → Fallback rule reassigns them to harm classes, lowering false negatives. &  0.355 & 0.365 & 0.256\\

08–10 & Added ten more examples and detailed, step-by-step formatting / labeling instructions for consistency. &  Inconsistent output format → Explicit formatting rules decrease false-positive and improve consistency.  & 0.369 & 0.362 & 0.301
\\
\grayrow
11–12 & Added two additional examples that blend profanity, sarcasm, and borderline harassment; guidance explains how to distinguish benign comments from mild harassment in such edge cases.   &  Profanity-sarcasm combinations mislabeled → Edge-case examples correct these errors.  & 0.382 & 0.357 & 0.282
 \\
13 & Expanded every definition with additional edge-case language (e.g., mixed sexist + homophobic slurs) and scenario lists; kept the same examples while refining dominant intent logic. & Overlapping harms confused the model→ Dominant-intent logic resolves conflicts. & 0.373 & 0.357 & 0.279 \\
\grayrow
14–16 & Added explicit rules for victim blaming, renaming that class "Deflection" after earlier prompts scored zero F1; updated the overlap-resolution hierarchy to include this new class.& Victim-blaming labeled “None” → New "Deflection" class increased recall and precision.  & 0.402 & 0.335 & 0.254
\\
17 & Rewrote "Discredit" and "Sexual Harassment" definitions with clearer behavioral triggers; kept the balanced set of 24 examples.& Definitions were too long and confusing → Shorter wording improved MCC, but sarcasm misclassifications returned. & 0.490 & 0.363 & 0.312 \\
\grayrow
18 & Reduced the examples to 18 for clarity; kept at least one example for every category and added three additional cases for Objectification, Maternal Insults, and Physical Appearance—categories that previously performed poorly. & Too many examples caused confusion → Fewer examples improved weaker categories, slight MCC changes across models. & 0.485
 & 0.398 & 0.383\\
19 & 33 carefully curated examples, detailed reasoning steps for sarcasm and tone, and a comprehensive overlap-resolution framework guiding label choice and confidence score. & Residual sarcasm and overlap errors → Comprehensive rules yield best precision–recall balance. & 0.501 & 0.451 & 0.362 \\
\hline
\end{tabular}
}
}

\vspace{-15pt}
\end{table*}

\begin{figure}[h]
    \centering
    \includegraphics[width=\linewidth]{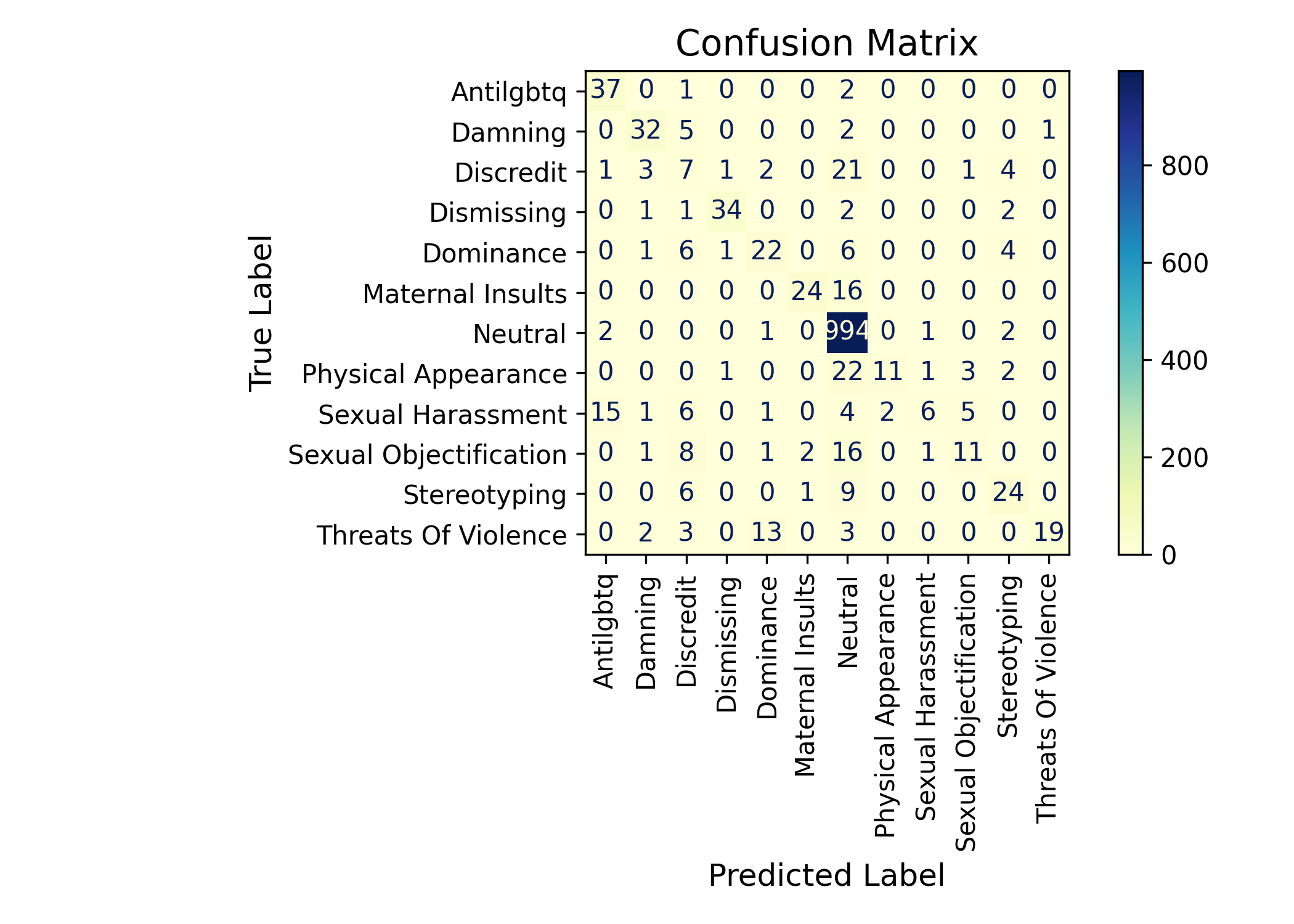}
    \caption{Confusion matrix for GPT-4o  using Prompt 19}
    \vspace{-15pt}
    \label{fig:confmatrix_gpt4o_phase3}
\end{figure}

\begin{boxedtext}
    \textit{Finding \#3:} Iterative refinement helped improve all models' performance over their baseline.  GPT-4o achieved the best MCC score (0.501) with  Prompt 19. 
\end{boxedtext}

\subsection{What are the strengths and limitations of the best model?}
To assess model performance across categories, we analyzed the confusion matrix from the final evaluation set (Figure~\ref{fig:confmatrix_gpt4o_phase3}), which revealed both the strengths and limitations of the best model. The model excelled in well-represented, distinct categories, correctly identifying 37 of 40 \textit{Anti-LGBTQ+} instances, 34 of 40 \textit{Dismissing} examples, and 994 \textit{Neutral} cases, the latter reflecting its bias toward the prevalent class. Conversely, it struggled with nuanced categories. For \textit{Discredit}, only 7 of 40 were correctly classified, with 21 mislabeled as \textit{Neutral}, conflating discrediting remarks with general feedback. Similarly, 15 \textit{Sexual Harassment } instances were mistaken for \textit{Anti-LGBTQ+}, and 16 Sexual Objectification examples were misclassified as \textit{Neutral}, indicating low recall and overlap in harm detection. These results underscore the model’s reliability for clear, frequent categories but highlight persistent challenges with subtle or overlapping harm, necessitating refined prompts, clearer label definitions, and enhanced training examples.

We also computed the binary performance of this model (i.e., whether it correctly identified a text as sexist or misogynistic if it belonged to one of its 11 forms. On those measures, it achieves  98.25\%  precision, 76.6\%  recall, and 86.1\% F1-score. These measures outperform the performance reported by SGI4SE~\cite{sultana2023automated}, which achieved 81.5\% precision, 70.3\% recall, and 75.3\% F1-score. However, we would like to mention that an apples-to-apples comparison is not feasible between our best model and SGID4SE, since the latter is trained on a limited set of sexist and misogynistic comments from the SGID dataset, whereas SOTA LLMs such as GPT-4o are trained on a large and diverse dataset.

\begin{boxedtext}
\textit{Finding \#4:} The best model performed well on frequent or distinct categories, such as Neutral, Anti-LGBTQ+, and Dismissing, achieving high F1-scores. However, it struggled significantly with nuanced categories, often misclassifying Discredit and Sexual Objectification as Neutral, and confusing Sexual Harassment with Anti-LGBTQ+.
\end{boxedtext}

\subsection{What are the most common misclassifications from the best-performing model?}

To better understand the model's behavior beyond standard accuracy metrics, we manually evaluated a subset of the model's predictions and their corresponding reasoning. In this analysis, the labels provided in the SGID dataset were treated as ground truth. We examined whether the model’s predicted categories matched the ground-truth labels and whether the accompanying explanation provided a valid justification for the prediction. This dual-layer assessment enabled us to identify key patterns of misclassification, including the failure to detect dismissive or sarcastic tone, confusion between overlapping categories, and difficulty in interpreting implicit harm. Although this evaluation involved some level of subjectivity, we adhered to the SGID rubric definitions throughout the process to maintain consistency and fairness. This analysis helped identify limitations in model reasoning, particularly in cases involving subtle, context-dependent instances.
This section examines common patterns of misclassifications observed across GPT-4o during the final evaluation phase. 

\vspace{2pt}
\noindent \emph{False positives:}
We identified only six instances of binary false positives. These comments were labeled as neutral, as they did not convey derogatory intent toward women or LGBTQ+ individuals, despite containing sexually explicit or potentially offensive keywords. For instance, the comment \textcolor{teal}{`\textit{`Lets just drop unauthorized erotic action on duty}"} includes sexual language but is not directed at any individual. Similarly, the comment \textcolor{teal}{``\textit{Shouldnt it be slave instead in this warning}"}, was flagged under the “dominance” category by GPT-4o due to the presence of the term “slave,” which can be associated with power dynamics. However, in this context, the comment appears to reference the “master-slave” software architecture, rather than expressing dominance over a person. This observation suggests that, when the target is not explicit, LLMs may mistakenly assume gendered identity and incorrectly mark as sexism or misogyny.

\vspace{2pt}
\noindent \emph{False negatives:} Despite advancements in definitions and the integration of reasoning capabilities, the model frequently failed to detect subtle or context-dependent instances of harm. We identified 103  binary false negative cases out of 440, which were further analyzed to determine the categories most frequently overlooked by GPT-4o. 

 The most commonly missed category was \emph{Physical Appearance}, accounting for 22 cases. In particular, the model failed to flag comments that referenced women’s clothing, such as \textcolor{teal}{``\textit{a joke for you Two dyslexics walk into a bra}''} and \textcolor{teal}{``\textit{Somewhat questionable that a skirt covers your legs}''}, both of which were labeled as neutral. These remarks, however, involve inappropriate references to women's attire that are not suitable for professional communication. Additional examples that were misclassified as neutral include comments targeting individuals’ physical features, such as \textcolor{teal}{``\textit{yo mama so fat I'm actually really concerned about her health}''} and \textcolor{teal}{``\textit{Not a stray curvy boiiiiiii}''}. The model justified its classification with statements like \textit{``expresses concern about health without mocking or insulting''} and \textit{``}playful remark without gendered or harmful context\textit{''}. However, these examples clearly exhibit body shaming and maternal insults, indicative of sexist undertones. In these cases, the model failed to recognize the nuanced expression of sexism.

 The second most frequently overlooked category was \emph{Discredit}, with 21 instances. This type of content often manifests in subtle forms, such as \textcolor{teal}{``a\textit{nother high school girl}''} or \textcolor{teal}{``added Mean Girls''}, which implicitly criticize someone's capabilities or dismiss women in a derogatory manner. GPT-4o misclassified these examples, reasoning that phrases like \textit{``}Mean Girls\textit{''} were neutral and lacked harmful language. However, such references are inappropriate in professional settings and convey implicit disrespect.

 The third most frequently missed category was \emph{Sexual Objectification}, where the model failed to flag comments containing pejorative language. The model justified these omissions by suggesting the remarks were not directed at a specific individual and instead expressed general frustration or profanity. For example, the comment \textcolor{teal}{``\textit{whys my shit on here cunt}''} was not flagged, despite containing explicit and offensive language. Clearly, the use of such terms should be considered sexist/misogynistic content regardless of whether a direct target is named.

\vspace{2pt}
\noindent \emph{Multiclass misclassifications:}
We further examined the patterns of misclassification across different sexism and misogyny categories. One of the most frequent types of confusion occurred between the categories of sexual harassment and anti-LGBTQ+. In these instances, comments were often misclassified as anti-LGBTQ+ solely due to the presence of keywords associated with LGBTQ+ identities. For example, \textcolor{teal}{\textit{``Pizza is not gay enough. Maybe more gay bacon strips}''} was classified as anti-LGBTQ+ based on keyword matching; however, according to the task definitions, such remarks that mock or express aggression based on sexuality should be categorized as sexual harassment. Additionally, we observed that 15 instances belonging to the sexual objectification category were incorrectly labeled as \textit{Discredit}. For instance, \textcolor{teal}{\textit{``Trash stupid Stupid cunt. This author is a stupid cunt I want to cry''}} was labeled as discredit, with the rationale that it attacks intelligence using gendered slurs. Nevertheless, the content involves the objectification of an individual. Moreover, a smaller number of comments originally labeled under the dominance category were frequently misclassified as \textit{Discredit}. For example, \textcolor{teal}{``\textit{Stop being so emotional about code reviews. This is why women don't belong in tech}''} demonstrates an exertion of authority intended to marginalize women, rather than criticizing any specific skills or competencies. These observations highlight the challenges in distinguishing subtle yet critical differences between sexism and misogyny categories.

 \begin{boxedtext}
\textit{Finding \#5:} Rare false positives occurred due to the lack of an explicit target. False negatives were more frequent and occurred due to GPT-4o's difficulty in reliably interpreting nuanced, context-dependent sexism and misogyny.
\end{boxedtext}

\section{Implications}
\label{sec:discussion}

This section summarizes key insights and failure patterns observed during model evaluation.

\vspace{3pt}
\noindent\emph{(Lesson \#1) Imbalanced datasets skew toward neutral predictions:} 
 In our initial exploratory phases, harmful and \emph{None} comments were equally represented. This caused models to over-predict harmful content, leading to false positives and poor generalization. For example, \textcolor{teal}{\textit{``the grammar on this is a bit funny maybe That every woman has the ability to code''}} was misclassified as Neutral comment. However, the statement promotes gender-based assumptions about women's coding abilities, making it a harmful example of stereotyping rather than a neutral remark. This highlights the need for better balancing and clearer differentiation between harmful and neutral categories. Data augmentation or hierarchical classification could balance predictions, reducing the risk of overlooking harmful content, a critical step for effective moderation.

\vspace{3pt}
\noindent\emph{(Lesson \#2)  Implicit harms require advanced contextual analysis:} 
Our model often missed sarcastic or subtle comments, like \textcolor{teal}{\textit{``nice try, leave coding to the pros''}}, which downplays women’s skills. It struggled with categories like \emph{Dismissing (F1 score: 0.75)} or \emph{Sexual Harassment (F1 score: 0.68)} because they blend technical talk with harm. Training models to understand comment threads or adding open-source context can help moderators catch these issues.

\vspace{3pt}
\noindent\emph{(Lesson \#3) Limitations in Handling Overlapping Categories:} 
 Misclassification frequently occurred between closely related categories such as \emph{Stereotyping} vs. \emph{Sexual Objectification}, or \emph{Discredit} vs. \emph{Damning}. For example, \textcolor{teal}{\textit{``Sort Clean Up, Halfling Female, Heads''}}, which discredits a contributor’s work but was labeled as \emph{Stereotyping} instead of \emph{Discredit}. The model saw it as a gender stereotype due to the female reference, missing the dismissive tone. These confusions suggest that category boundaries may be too ambiguous for current models. Future work may benefit from hierarchical classification or clearer example-driven boundary definitions.

\vspace{3pt}
\noindent\emph{(Lesson \#4): Models overgeneralize to Neutral when harm is implicit: }

Comments involving sarcasm, vague insults, or playful tone were frequently mislabeled as \emph{Neutral}. For example, the statement \textcolor{teal}{\textit{``You too can be lesbian''}} was classified as \emph{Neutral}, despite its dismissive and sexually suggestive framing. This result suggests a need for models that more effectively capture subtext and social nuance.

\vspace{3pt}
\noindent\emph{Lesson \#5: Label phrasing directly affects model comprehension.} 
The original label \emph{Victim Blaming} was too abstract and required understanding intent, making it difficult for models to detect. We first renamed it to \emph{Deflection}, which focused on shifting blame or responsibility. This wording helped clarify the behavior, and for the first time, models like GPT-4o correctly identified a few examples.  
However, performance was still limited, so we refined the label further to \emph{Dismissing}, which emphasized observable actions like downplaying, minimizing, or silencing someone’s experience. This final phrasing significantly improved detection across models, resulting in 30 true positives in the final evaluation.
These findings collectively underscore the importance of domain-specific tuning and evaluation in enhancing the reliability of LLM-based moderation systems.

\vspace{3pt}
\noindent\emph{(Lesson \#6) Nuance Detection Remains a Significant Challenge:} Even after extensive prompt optimization, LLMs still struggle with the most subtle, context-dependent, or overlapping forms of harm. Specific confusion patterns persist (e.g., mistaking subtle harm for neutral text, conflating different types of hostility), indicating inherent limitations in current model capabilities or prompting techniques for these edge cases.

\vspace{3pt}\noindent\emph{(Lesson \#7) Granular Evaluation Reveals Key Insights:} Relying solely on overall metrics is misleading. Analyzing performance per category (using F1-scores and confusion matrices) and employing metrics robust to imbalance (such as MCC) are essential to understanding precisely where models succeed and fail, thereby guiding refinement efforts effectively.

\vspace{3pt}\noindent \emph{(Lesson \#8)  Reliable Binary Precision: } While the best model had $\approx$24\% false negatives, false positives were very low 1.7\%. This result suggests a promising application of prompt-tuned LLMs, particularly when the goal is binary flagging.

\section{Threats to Validity}
\label{sec:threats-to-validity}
\textbf{Internal Validity:} Our selection of the SGID dataset developed by Sultana \textit{et} al. for this study is a threat to internal validity.   While labels were grounded in real examples and definitions, some categories may be missing due to uneven representation across projects or their sampling criteria. We used the original labels without modification and ensured category balance wherever possible.

\vspace{2pt}
\noindent \textbf{Construct Validity:} Label definitions were refined during prompt development to improve clarity. While we refined category definitions (e.g., renaming \texttt{Victim Blaming} to \texttt{Dismissing} and aligning them with empirical examples during prompt engineering, subjective interpretations of harm can still vary across annotators or cultural contexts. Additionally, model classifications rely heavily on linguistic patterns, which may not fully capture sociocultural nuance or intent. To minimize this threat, we included fine-grained definitions, a representative few-shot example, and cross-model comparisons during evaluation. While we evaluated 3 SOTA models using 20 iteratively refined prompts, even further optimization may be possible. Regardless, our study provides valuable guidelines for building an SE domain-specific multiclass classifier and establishes a baseline in sexism/misogyny classification. 

While we experimented with prompt refinement and parameter tuning, we were unable to fine-tune the LLM weights, as this requires a large-scale labeled dataset, which is currently unavailable.

\vspace{2pt}
\noindent \textbf{External Validity:} Our experiments are based entirely on GitHub comments and may not generalize to other platforms with different language norms, such as Reddit or Stack Overflow. While the classification framework is adaptable, prompt and schema adjustments would be needed for other domains. Moreover, sexism and misogyny are subjective and depend on various demographic factors. As definitions used in our prompt are not generalizable, neither are classifiers built using these prompts.


\section{Related Works}
\label{sec:background}

\emph{Misogynistic content identification:}
Iberval introduced two annotated datasets consisting of English and Spanish tweets and organized the Automatic Misogyny Identification (AMI) task in 2018~\cite{fersini2018overview}, which attracted considerable interest from researchers and practitioners aiming to develop systems for detecting misogynistic content. In addition to the datasets, they provided a rubric for classifying misogynistic text into six categories: (i) Stereotyping and Objectification, (ii) Dominance, (iii) Derailing, (iv) Sexual Harassment and Threats of Violence, (v) Discredit, and (vi) Non-misogynistic~\cite{fersini2018overview}. Motivated by this initiative, EVALITA hosted two AMI competitions in 2019 and 2020, focusing on Italian-language tweets~\cite{fersini2018overview,fersini2020ami}. In 2020, Muti et al.\ achieved the highest reported F1-score of 74.3\% using the AlBERTo model~\cite{muti2020unibo} for the AMI task. Subsequent research has extended misogyny detection to multiple online platforms, including YouTube comments~\cite{shila_1}, Reddit posts~\cite{Farrell}, and Twitter posts~\cite{jha-mamidi-2017-compliment,guest2021expert}. Moreover, the focus has expanded from textual content to multimodal data, with growing interest in detecting misogyny within memes through deep learning-based approaches. EXIST 2024 has hosted a task for misogynous meme identification, sharing a dataset of memes curated from Google~\cite {plaza2024overview}. This competition has resulted in building numerous multimodal approaches~\cite {fersini2019detecting,maqbool2024contrastive} for such content identification, where mDeBERTa-v3 has outperformed all regarding meme identification~\cite{ruiz2024concatenated} with a 76.4\% F1 score.

\emph{Automated content moderation using LLM:} The proliferation of user-generated content on digital platforms has necessitated robust content moderation strategies to mitigate the spread of hate-based, abusive, or inappropriate material. Traditional moderation approaches, often reliant on rule-based systems and keyword filtering, have struggled to address the nuanced and context-dependent nature of online content. The advent of large language models (LLMs), such as GPT-3, GPT-4, LLaMA, and Gemini, has introduced new possibilities for automated content moderation due to their advanced natural language understanding capabilities. Recent studies have demonstrated that LLMs can effectively perform tasks such as toxic content detection and rule-based community moderation, outperforming traditional classifiers in various benchmarks~\cite{aldahoul2024advancing,kumar2024watch}.
Using a dataset of gender-based violence (GBV) tweets, Aldahoul et al.~\cite{aldahoul2024advancing} evaluated large language models (LLMs) for zero-shot content moderation. They found that GPT-4o and GPT-4o-mini outperformed Gemini and the OpenAI moderation model, achieving the lowest false negative rates in detecting GBV.
Kumar et al.~\cite{kumar2024watch} used a zero-shot approach on a balanced Reddit dataset of norm violations and found GPT-3.5 effectively applied subreddit-specific rules, with 64\% median accuracy and 83\% precision. 

Researchers have also explored few-shot prompting to enhance LLM performance on nuanced moderation tasks.
Bonagiri et al.~\cite{bonagiri2025towards} demonstrated that few-shot learning enables LLMs to outperform standard classifiers in identifying harmful content, with GPT-4o-mini achieving up to an 81\% F1-score on a dataset of annotated YouTube videos. Using the same dataset, Oak et al.~\cite{oak2025re} developed a re-classification framework where GPT-3.5 consistently surpassed models like Mistral-7B and LLaMA2-13B.

\emph{Derogatory content identification for software developers:} Researchers have explored the prevalence of toxic and uncivil language, as well as user “pushback,” within open-source developer forums and platforms~\cite{raman2020stress,miller2022did,sarker-tosem-toxicr,sarker2023toxispanse}. A wide range of machine learning techniques have been applied to identify toxic content, ranging from traditional models like Support Vector Machines (SVM)~\cite{raman2020stress} to transformer-based architectures~\cite{sarker-tosem-toxicr,sultana2023automated}. However, research specifically focused on detecting misogynistic or gender-based derogatory content in these environments remains in its early stages, despite numerous documented instances of such discrimination~\cite{singh2021motivated}. Such content often span a nuanced spectrum that differs from conventional hate speech, trolling, or explicit violations of community guidelines~\cite{sultana2023automated}. Prior research has also shown that the characteristics of derogatory content in software development communities are qualitatively different from similar content found in broader online platforms~\cite{sultana2023automated,squire2015floss}. These contextual differences underscore the need for specialized moderation systems tailored to the unique communication dynamics of open-source software (OSS) communities. With the increasing application of large language models (LLMs) in software engineering tasks, recent studies have begun to investigate their potential in identifying toxic and gender-biased language in developer communications~\cite{mishra2024exploring,sultana2024exploring}. Therefore, we investigate the potential of large language models (LLMs) to detect such forms of derogatory content within open-source software (OSS) communities. Our results show that LLMs perform better than traditional transformer-based models, especially when it comes to fine-grained classification of nuanced and context-sensitive language~\cite{sultana2023automated}.

\section{Conclusion}
\label{sec:conclusion}

This study evaluated the effectiveness of large language models (LLMs) in detecting sexist and harmful content in GitHub comments. Using prompt engineering across 12 fine-grained categories, we evaluated multiple LLMs and refined the prompts to enhance multiclass classification performance. Although the best model had high binary precision, its binary recall suffered due to challenges in identifying subtle or context-dependent harm, such as sarcasm or overlapping categories like \emph{Discredit}, \emph{Damning}, and \emph{Dismissing}.

In terms of multiclass classification, the best model accurately labeled more than 60\% instances belonging to \emph{Anti-LGBTQ+}, \emph{Maternal Insults}, \emph{Dismissing},  \emph{Damning}, \emph{Stereotyping}, and \emph{Neutral}. However, it misclassified more than 70\%  of the instances belonging to  \emph{Discredit}, \emph{Physical appearance}, and \emph{Sexual harassment}, and \emph{Sexual objectification}. 
Our evaluation also demonstrates the need for prompt refinement, inclusion of conversational context, more explicit category definitions, and expert review of edge cases.

\bibliographystyle{IEEEtran}
\bibliography{references}


\end{document}